\documentclass[aps,prb,twocolumn,showpacs,superscriptaddress,floatfix]{revtex4-2}

\usepackage{graphicx}
\usepackage{amsmath,amssymb,amsfonts}

\begin{document}

\title{Enhancement of density of states and suppression of superconductivity in site-disordered topological metal LaPtSi}



\author{Sitaram Ramakrishnan}
\email{niranj002@gmail.com}
\affiliation{Department of Quantum Matter, AdSE,
Hiroshima University, Higashi-Hiroshima 739-8530, Japan}

\author{Tatsuya Yamakawa}
\affiliation{Department of Quantum Matter, AdSE,
Hiroshima University, Higashi-Hiroshima 739-8530, Japan}

\author{Ryohei Oishi}
\affiliation{Department of Quantum Matter, AdSE,
Hiroshima University, Higashi-Hiroshima 739-8530, Japan}

\author{Yasuyuki Shimura}
\affiliation{Department of Quantum Matter, AdSE,
Hiroshima University, Higashi-Hiroshima 739-8530, Japan}

\author{Takahiro Onimaru}
\affiliation{Department of Quantum Matter, AdSE,
Hiroshima University, Higashi-Hiroshima 739-8530, Japan}

\author{Arumugam Thamizhavel}
\email{thamizh@tifr.res.in}
\affiliation{Department of Condensed Matter Physics
and Materials Science,
Tata Institute of Fundamental Research, Mumbai 400005, India}

\author{Srinivasan Ramakrishnan}
\email{ramky07@gmail.com}
\affiliation{Department of Physics, Indian Institute of Science
Education and Research, Pune, 411008, India}

\author{Minoru Nohara}
\email{mnohara@hiroshima-u.ac.jp}
\affiliation{Department of Quantum Matter, AdSE,
Hiroshima University, Higashi-Hiroshima 739-8530, Japan}

\date{\today}

\begin{abstract}
Single crystals of non-centrosymmetric $s$-wave superconductor LaPt$_{0.88}$Si$_{1.12}$ have been grown by the Czochralski (Cz) technique,
whose crystal structure is described by the space group $I4{_1}md$  at ambient
conditions. The inter-site mixing between platinum and silicon is confirmed by both single-crystal x-ray diffraction (SXRD) and
electron probe micro-analyzer (EPMA). The disordered material exhibits a lower superconducting (SC) transition
temperature $T_c$ at 2.02 K as opposed to
the highest value of 3.9 K reported in polycrystalline LaPtSi without inter-site mixing.
From specific heat, the Sommerfeld coefficient
($\gamma$) is estimated to be 7.85  mJ/mol K$^2$, which is much larger than the values reported for the samples
exhibiting higher $T_c$.
This is unprecedented as $T_c$ seems to decrease with increase in
the electron density of states (DOS) at the Fermi energy and thus $\gamma$.
The present work reports on the anomalous behaviour of SC and normal state properties of
LaPt$_{x}$Si$_{2-x}$, presumably caused due to the existence of non-trivial topological bands.

\end{abstract}

\maketitle

\clearpage

\section{\label{sec:Laptsi_introduction}Introduction}
Superconductivity (SC) in non-centrosymmetric
compounds have been of significant interest for condensed matter
physicists for almost two decades ever since the discovery of
the heavy fermion superconductor CePt$_3$Si \cite{bauer2004a}.
Breaking of center of inversion in superconductors leads
to promotion of exotic phenomena like spin singlet and triplet mixing
in the superconducting phase
due to antisymmetric spin orbit coupling (ASOC) \cite{gorkov2001a, yuan2006a, bauer2009a, smidan2017a}.

One such compound is LaPtSi whose
non-centrosymmetric tetragonal crystal structure ($I4{_1}md$) was initially reported
by Klepp \textit{et al.} employing powder x-ray diffraction (PXRD) \cite{klepp1982a}. Bulk SC was discovered
at $T_c$ = 3.3 K by Evers \textit{et al.} through ac susceptibility measurements \cite{evers1984a}.
Ramakrishnan \textit{et al.} \cite{ramakrishnan1995a} revealed that it is
fully gapped, weakly coupled dirty $s$-wave superconductor where the $T_c$
was comparatively higher at 3.9 K than the value reported by the earlier study \cite{evers1984a}. Later,
Kneidinger \textit{et al.} \cite{kneidinger2013a} observed
a similar behaviour in the SC of LaPtSi with a $T_c$ = 3.35 K
close to that of Evers \textit{et al.} \cite{evers1984a}.
Previous work \cite{kneidinger2013a} has suggested that superconductivity in LaPtSi may originate predominantly
from Pt and La electronic states since Pt-d and La-d states
dominate the electronic density of states (DOS) at the Fermi energy.
Further, they have concluded that despite the  large spin-orbit coupling (SOC) in LaPtSi,
the topology of the Fermi surface probably enforces predominantly spin-singlet pairing.
They claimed that this could explain the absence of the  unconventional SC in
LaPtSi. Magnetic-penetration-depth measurements down to 0.02 $T_c$ on LaPtSi confirm
that it is fully gapped and cannot be considered as a
candidate as a time-reversal symmetric breaking topological superconductor
despite the presence of a strong ASOC \cite{palazzese2018a}.

Interest in LaPtSi has been revived recently as first-principles calculations suggest
that the topology of the bands is non-trivial in both LaPtSi and LaPtGe, where
both are Dirac superconducting semimetals \cite{zhang2020a, shi2021a}.
They report that the presence of non-trivial topological $Z_2$
invariants and surface states on the (001) surface provide a convenient test bed for the study
of the interplay of bulk Dirac points and superconductivity. Earlier studies
\cite{klepp1982a, ramakrishnan1995a, kneidinger2013a, palazzese2018a} have failed to observe this possibly due to lattice disorder
as all experimental investigations have been done on polycrystalline
material of LaPtSi. In fact, a recent
theoretical study \cite{mahdi2023a} suggests that the disorder strongly affects the topological phase by
closing the energy gap, while trivial SC phases remain stable and fully gapped.

In the present work, we grew a single crystal by the Czochralski (Cz) method resulting in LaPt$_{0.88}$Si$_{1.12}$ confirmed
by single-crystal x-ray diffraction (SXRD) and
electron probe micro-analyzer (EPMA), where it is observed that the inter-site mixing exhibits
a significant decrease in $T_c$
from 3.9 K \cite{ramakrishnan1995a}
down to 2.02 K.
Moreover, despite the reduction of $T_c$ compared to stoichiometric LaPtSi, we see
an enhancement of the Sommerfeld coefficient ($\gamma$) which is unusual.
Ideally, when the DOS is reduced, the $T_c$ along with $\gamma$ should have been smaller.
We propose that the present observed effects on the normal and SC state
properties of LaPt$_x$Si$_{2-x}$ possibly stems from the underlying nature of the non-trivial topology
of the crystal suggested by the theories  \cite{zhang2020a, shi2021a}.

\section{\label{sec:Laptsi_experimental}Experimental}
\subsection{Crystal growth optimization and Laue}

\begin{figure}[ht]
\includegraphics[width=80mm]{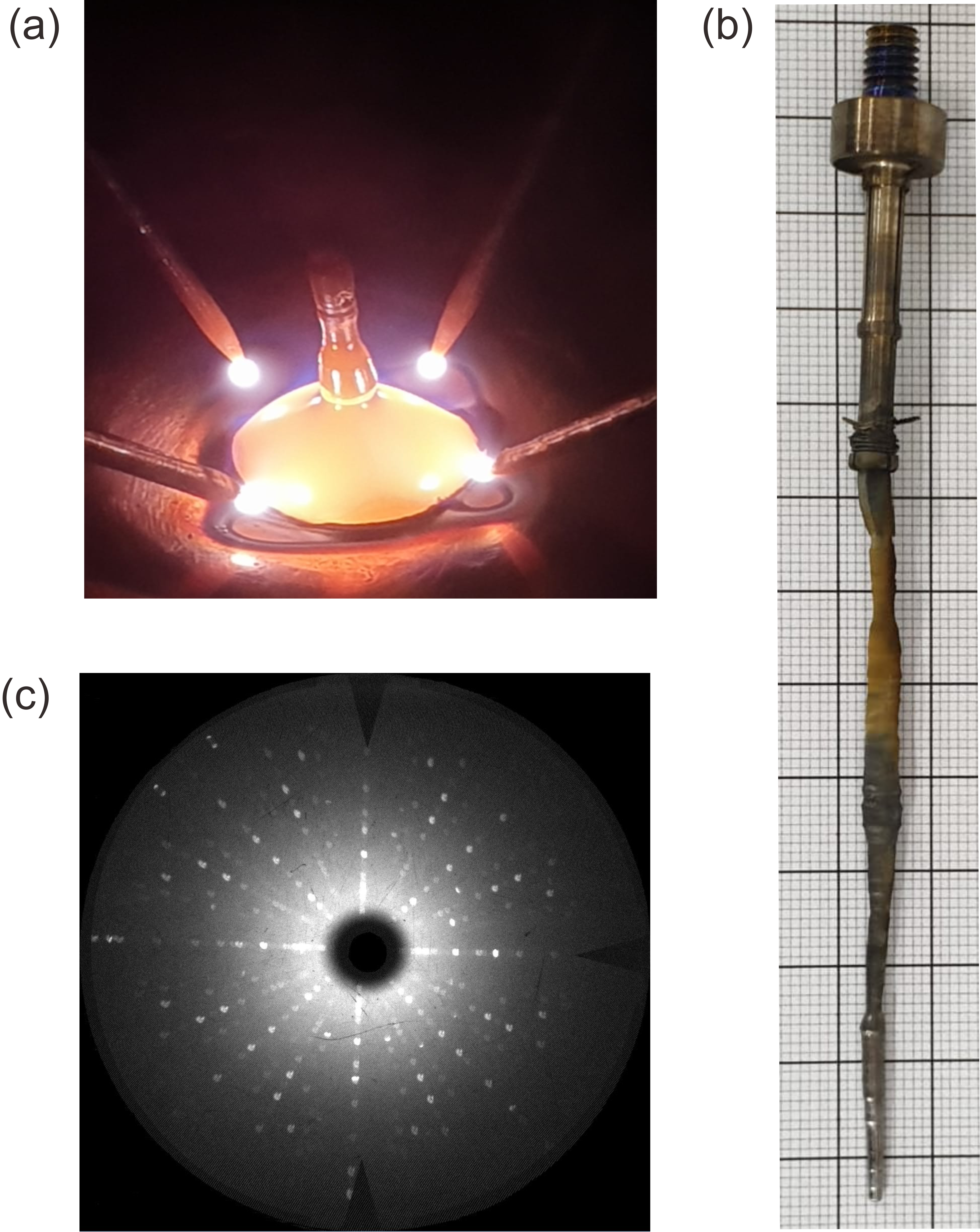}
 \caption{(a) Crystal growth in a tetra arc furnace by the Czochralski method. (b) A cylindrical shaped ingot
 was obtained of about 70 mm in length and 4 mm in diameter. (c) Laue pattern corresponding to (001) plane suggesting a good quality single crystal.}
  \label{crystalgrowth}
\end{figure}
The growth of the material was done by the Czochralski (Cz) method in a tetra-arc furnace (Technosearch Corporation, Japan) under an ultra-pure argon atmosphere in
an iterative process as shown in Figure \ref{crystalgrowth} (a).
The Cz method was chosen over flux growth or any other technique because
the individual elements have high melting points.  Initially, high purity
individual elements of La : Pt : Si (99.99\% for La and Pt, and 99.999\% for Si) were taken in the stoichiometric ratio 1 : 1 : 1, amounting to 10 g,
and melted repeatedly to ensure its homogeneity after which a polcrystalline seed crystal was cut from this
ingot for the purpose of crystal growth. The polycrystalline seed was gently inserted into the molten solution
and initially pulled at a rapid speed of about 90 mm/h.  The temperature of the melt was adjusted such
that a necking is formed and then we employed a pulling speed of about 10 mm/h throughout the growth process.
A 70 mm long ingot was pulled with a diameter of 3–4 mm. The
grown ingot was subjected to Laue diffraction where the diffraction spots were
 of rather poor quality. Hence,
the entire process was repeated with the same parameters using the first
pulled ingot as a seed to grow a second new single crystal (Figure. \ref{crystalgrowth} (b))
after which Laue diffraction confirmed the new crystal shows
good single crystallinity.
as seen in Figure \ref{crystalgrowth} (c).

\subsection{Electron probe micro-analysis}
The crystal was cut perpendicular to $c$ direction into several pieces by using
a wire electric discharge machine.
The composition of the pieces was checked by Electron probe micro-analyzer (EPMA) (JXA-iSP 100 Super probe)
with accelerating voltage of 20 kV and a beam current of 30 nA.
Data were collected on different crystals on a smooth surface after polishing.
The composition averaged from all the data is LaPt$_{0.90(3)}$Si$_{1.10(3)}$
where can one see that it is slightly Pt deficient with a subsequent increase in Si.

\subsection{Single-crystal x-ray diffraction}
Single-crystal x-ray diffraction (SXRD) was measured on a
four-circle Bruker diffractometer employing Mo K$\alpha$ radiation.
SXRD data were processed by the APEX-III software \cite{apex3} and structure refinements
were done using Jana 2006 \cite{petricekv2014a, petricekv2016a}. For the crystallographic table and further details
of the SXRD data collection refer to the supplemental material \cite{laptsisuppmat2023a}. The results from SXRD
are consistent with the composition determined by EPMA as
discussed in Section \ref{subec:Laptsi_SXRD} .

\subsection{Physical properties}
A commercial superconducting quantum interference device (SQUID) magnetometer (MPMS5,
Quantum Design, USA) was used to measure magnetic moment $M$ in a field
of 10 Oe as a function of temperature from 1.8 to
4 K to detect the superconducting transition and at 5 T to measure the normal state
susceptibility $\chi$ from 2 to 300 K.
The electrical resistivity between 1.8 and 300 K was
measured by the standard dc four probe technique in
a commercial Physical Property Measurement System (PPMS, Quantum
Design, USA).  The current was applied parallel to \textbf{a} axis,
and the magnetic field was applied perpendicular to it.
The superconducting transition temperatures under
different fields from 0 to 0.25 T were measured.
The specific heat measurements were carried out in the PPMS with
a $^3$He probe and data was collected
from temperatures from 0.4 to 15 K.

\section{\label{sec:Laptsi_results_discussion}%
Results and discussion}

\subsection{\label{subec:Laptsi_SXRD}
Evidence of Pt/Si site disorder by single-crystal x-ray diffraction}
\begin{figure}[ht]
\includegraphics[width=80mm]{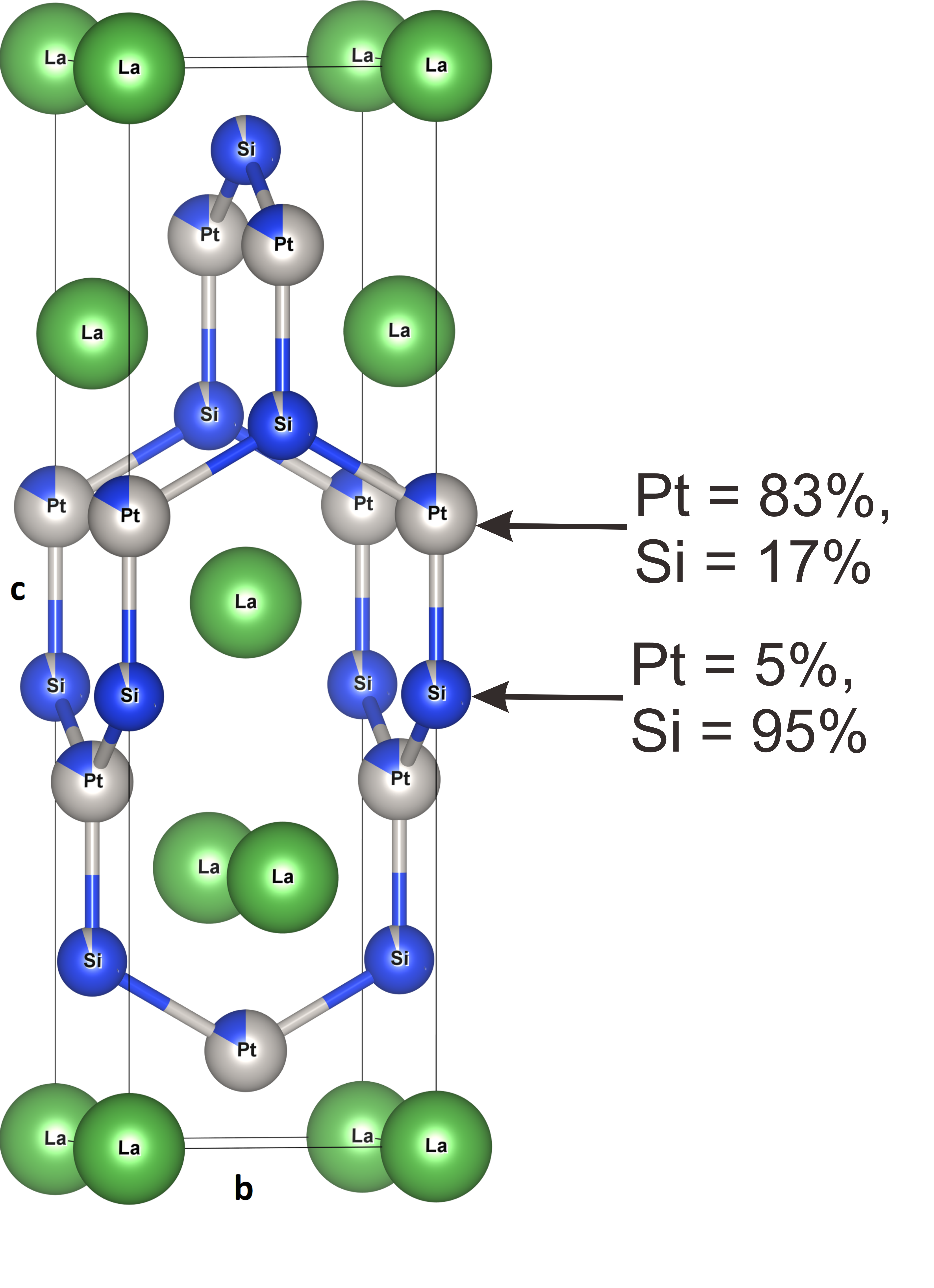}
 \caption{The non-centrosymmetric $I$-centered cell of LaPt$_{0.88}$Si$_{1.12}$ featuring
 the hyper-honeycomb network of platinum and silicon atoms. The large green colored spheres are lanthanum, blue
 and grey colored spheres of intermediate sizes correspond to silicon and platinum. For the platinum and silicon atomic sites one can see a mixture of both
 blue and grey colors indicative of Pt/Si site mixing. Arrows depict the percentage of inter-site mixing between Pt and Si.}
  \label{cell}
\end{figure}

As mentioned in sections II B and C, the composition of the material grown was determined
by EPMA and SXRD measurements.
All diffraction maxima could be indexed by a single unit cell with lattice parameters
$a = 4.2441(2)$ \AA{},  $c = 14.5264(2)$ \AA{}, similar to the published unit
cell of LaPtSi \cite{klepp1982a, ramakrishnan1995a, kneidinger2013a, palazzese2018a},
albeit with a slight reduction in volume as seen in Table \ref{comparelattice}. Figure \ref{cell} shows
the crystal structure of LaPt$_{0.88}$Si$_{1.12}$ which is
isostructural to LaPtSi where both are derivatives of the $\alpha$-ThSi$_2$ type
structure like LaSi$_2$ \cite{satoh1970a} or SrGe$_2$ \cite{akira2017a},
as the latter compounds are centrosymmetric
($I4{_1}$/${amd}$) where the Si or Ge atoms form a hyper-honeycomb
network in an $I$-centered lattice. In both LaPtSi and LaPt$_{0.88}$Si$_{1.12}$, 50\%
of the Si atoms are replaced by Pt. Thus, the hyper-honeycomb
network is formed by alternating Pt and Si atoms thereby
breaking the center of inversion, crystallizing
into the non-centrosymmetric $I4{_1}md$ which is a subgroup of $I4{_1}$/${amd}$.
LaPt$_{0.88}$Si$_{1.12}$ has inter-site mixing between Pt and Si
as seen in Figure \ref{cell}.
\begin{table*}[ht]
\small
\centering
\caption{Comparison of lattice parameters
of LaPt$_x$Si$_{2-x}$ ($x$ = 1 and 0.88) where
one can see a slight reduction in the volume for $x = 0.88$ due to Pt/Si site mixing.}
\begin{ruledtabular}
\begin{tabular}{ccccc}
  Type   & Polycrystal \cite{klepp1982a} &  Polycrystal \cite{ramakrishnan1995a} & Polycrystal \cite{kneidinger2013a} & Single crystal [Present work] \\
$x$  &    1  & 1  & 1  & 0.88 \\
Space group & $I4{_1}md$        & $I4{_1}md$             & $I4{_1}md$                   & $I4{_1}md$                        \\
$a$ (\AA)   &  4.2490(3)        & 4.2494(4)            & 4.2502(1)                    & 4.2441(2)                    \\
$c$ (\AA)  & 14.539(2)          & 14.5338(4)           & 14.5250(5)                    & 14.5264(2)                    \\
$V$ (\AA${^3}$) & 262.49(2)       & 262.44 (4)           & 262.38 (1)                   & 261.66 (2)                     \\
\end{tabular}
\end{ruledtabular}
\label{comparelattice}
\end{table*}

Table \ref{comparemodel} shows the structural analysis for two different models
on basis of the occupancy of Pt and Si to show Model II with Pt/Si disorder yields a better fit.

\begin{table}[ht]
\centering
\caption{\label{comparemodel}Comparison of two structure refinements based on the composition
of the compound upon consideration of Pt/Si disorder.}
\begin{ruledtabular}
\begin{tabular}{ccccccc}
Model               &I      & II                                                                   \\
\hline
Composition         & LaPtSi      & LaPt$_{0.88(3)}$Si$_{1.12(3)}$     \\
Space group         & $I4_{1}md$      & $I4_{1}md$     \\
Occ[La]             &  1     & 1           \\
Occ[Pt]             &  1     & 0.83(3)           \\
Occ[Si$_{Pt}$]      &  -     & 0.17           \\
Occ[Si]             &  1     & 0.95(3)           \\
Occ[Pt$_{Si}$]      &  -     & 0.05            \\
Unique reflections (obs/all) &  103/119     & 103/119            \\
No. of parameters   & 13      &   15            \\
$R_{F}$(obs)        & 0.0353      & 0.0311             \\
$wR_{F}$(all)       & 0.0555      & 0.0509              \\
GoF (obs/all)       & 3.97/3.77      &  3.67/3.49           \\
$\Delta\rho_{min}$, $\Delta\rho_{max}$($e$ \AA$^{-3})$ & -6.10, 6.24      & -3.95, 5.86      \\
\end{tabular}
\end{ruledtabular}
\end{table}

Initally, model I was tested where all atomic sites were completely filled.
The composition is LaPtSi and it leads to a reasonable fit of the
diffraction data with $R = 0.0353$. However, the atomic displacement parameters (ADP) for all atoms
are non-positive definite which is physically meaningless in the structure as
shown in Table S2 in the supplemental material \cite{laptsisuppmat2023a}.
In model II a silicon atom (Si$_{Pt}$) was introduced
on the Pt site and the occupancies of Pt and Si$_{Pt}$ were refined while keeping
the sum of the occupancies to be 1, subsequently a platinum atom was (Pt$_{Si}$)
was introduced on the Si site and their occupancies were refined while constraining
the sum to be 1, resulting in an improved fit
towards $R = 0.0311$ as seen in Table \ref{comparemodel}.
Moreover, the ADP of all atoms is now positive definite.
Table \ref{atom} shows the atomic coordinates and atomic
displacement parameters (ADPs) of Model II.

\begin{table*}[ht]
\centering
\small
\caption{\label{atom}%
Structural parameters for the tetragonal
structure of crystal A of LaPt$_{0.88}$Si$_{1.12}$  at 300 K.
Given are the fractional coordinates $x$, $y$, $z$ of the atoms,
their anisotropic displacement parameters (ADPs)
$U_{ij}$ $(i, j = 1, 2, 3)$
and the equivalent isotropic displacement parameter $U^{eq}_{iso}$.
$R_F$(obs/all)=0.0311/0.0363, no. of parameters is 15.
Refinement method used: least-squares on $F$. Space group: $I4_{1}md$.
Criterion of observability: $I < 3\sigma(I)$.}
\begin{ruledtabular}
\begin{tabular}{cccccccccccc}
Atom & Occ  & $x$ & $y$ & $z$ & $U_{11}$ & $U_{22}$ & $U_{33}$ & $U_{12}$ & $U_{13}$ & $U_{23}$ & $U^{eq}_{iso}$ \\
\hline
La &1 &0 & 0 & 0 &  0.0086(18) &   0.0050(19) &  0.0098(19) &  0 & 0 & 0 &   0.0078(11) \\
Pt &0.83(3) &0  &0 &0.5852(2) &  0.0108(10) &   0.0010(10) &  0.0042(9) &   0 & 0 & 0 &  0.0053(6) \\
Si$_{Pt}$ &0.17 &0  &0 &0.5852 &  0.0108 &   0.0010 &  0.0042 &   0 & 0 & 0 &  0.0053 \\
Si &0.95(3) &0 &0 &0.4188(7) &   0.0141(63) & 0.0093(67) &  0.0082(57) &0 &0 &0 &   0.0105(36) \\
Pt$_{Si}$ & 0.05 &0 &0 &0.4188 &   0.0141 & 0.0093 &  0.0082 &0 &0 &0 &   0.0105 \\
\end{tabular}
\end{ruledtabular}
\end{table*}

\subsection{Electrical resistivity}
\begin{figure*}[ht]
\centering
\includegraphics[width=80mm]{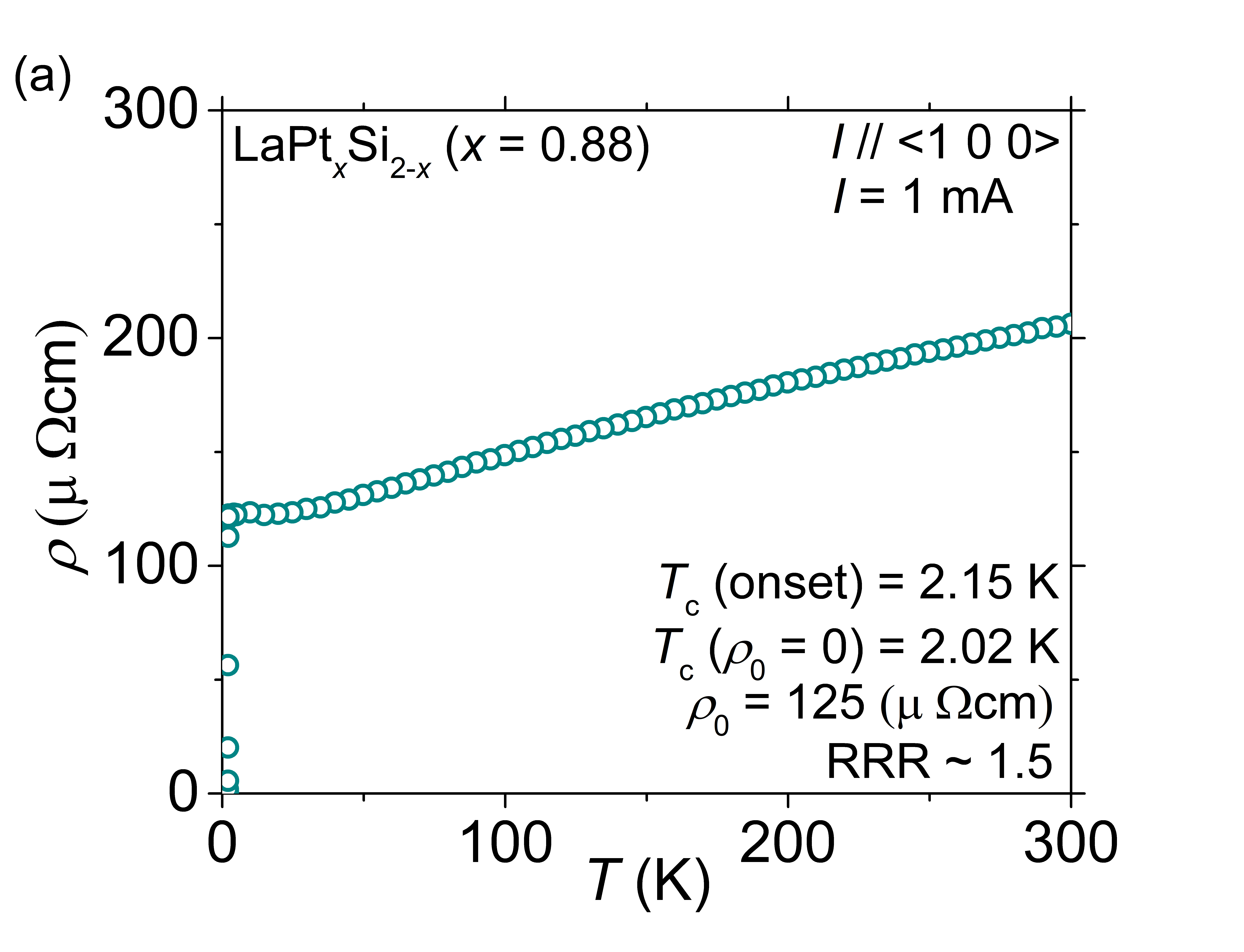}%
\includegraphics[width=80mm]{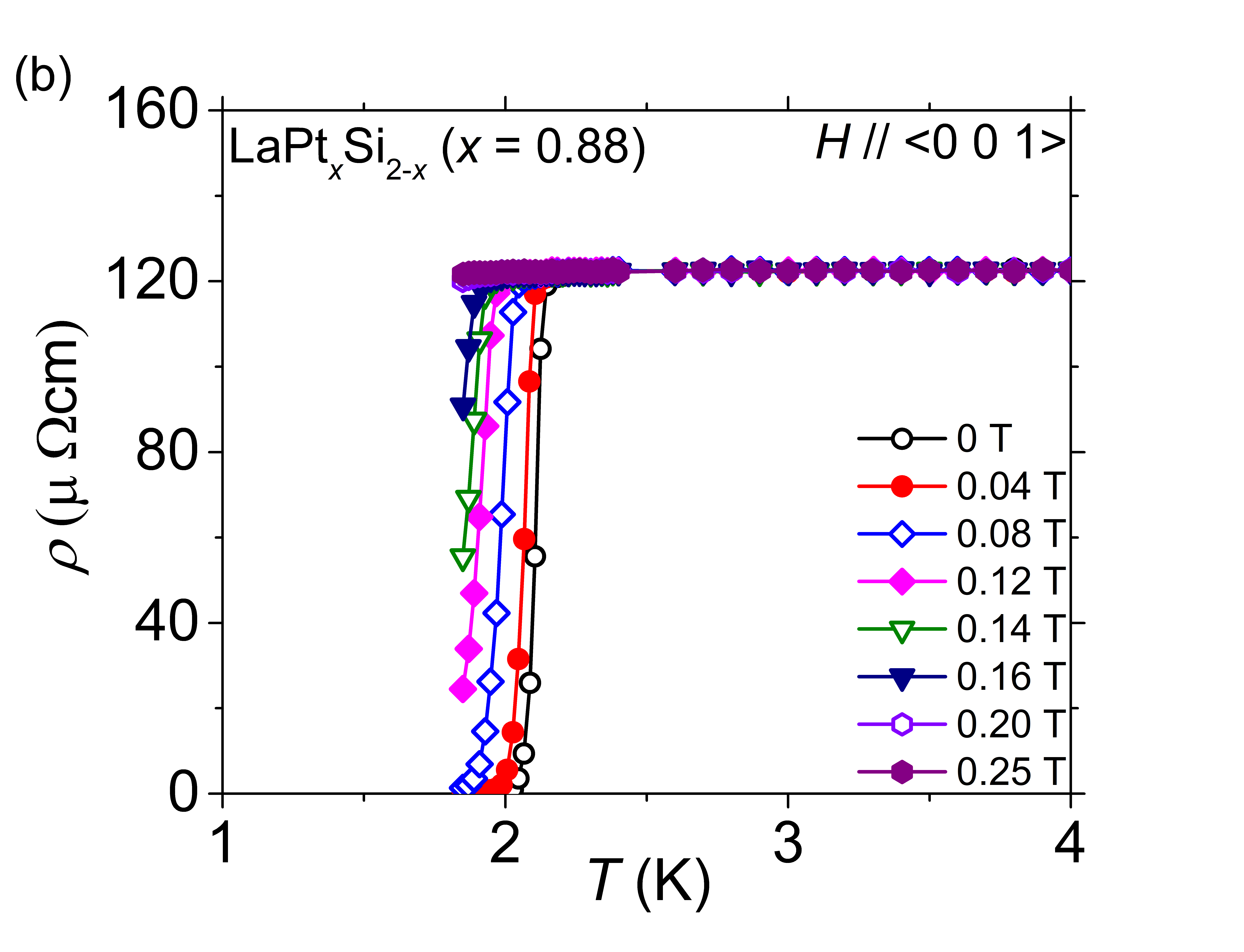}%
\caption{\label{fig:laptsi_resi}(a) Temperature dependence of the resistivity of LaPt$_{0.88}$Si$_{1.12}$ from 1.8 to 300 K. A current of 1 mA is applied parallel to the $a$ direction (b) Field-dependent
resistivity  measurements from 0 to 0.25 T where field is applied parallel to $c$ direction.}
\end{figure*}
\begin{figure}[ht]
\centering
\includegraphics[width=80mm]{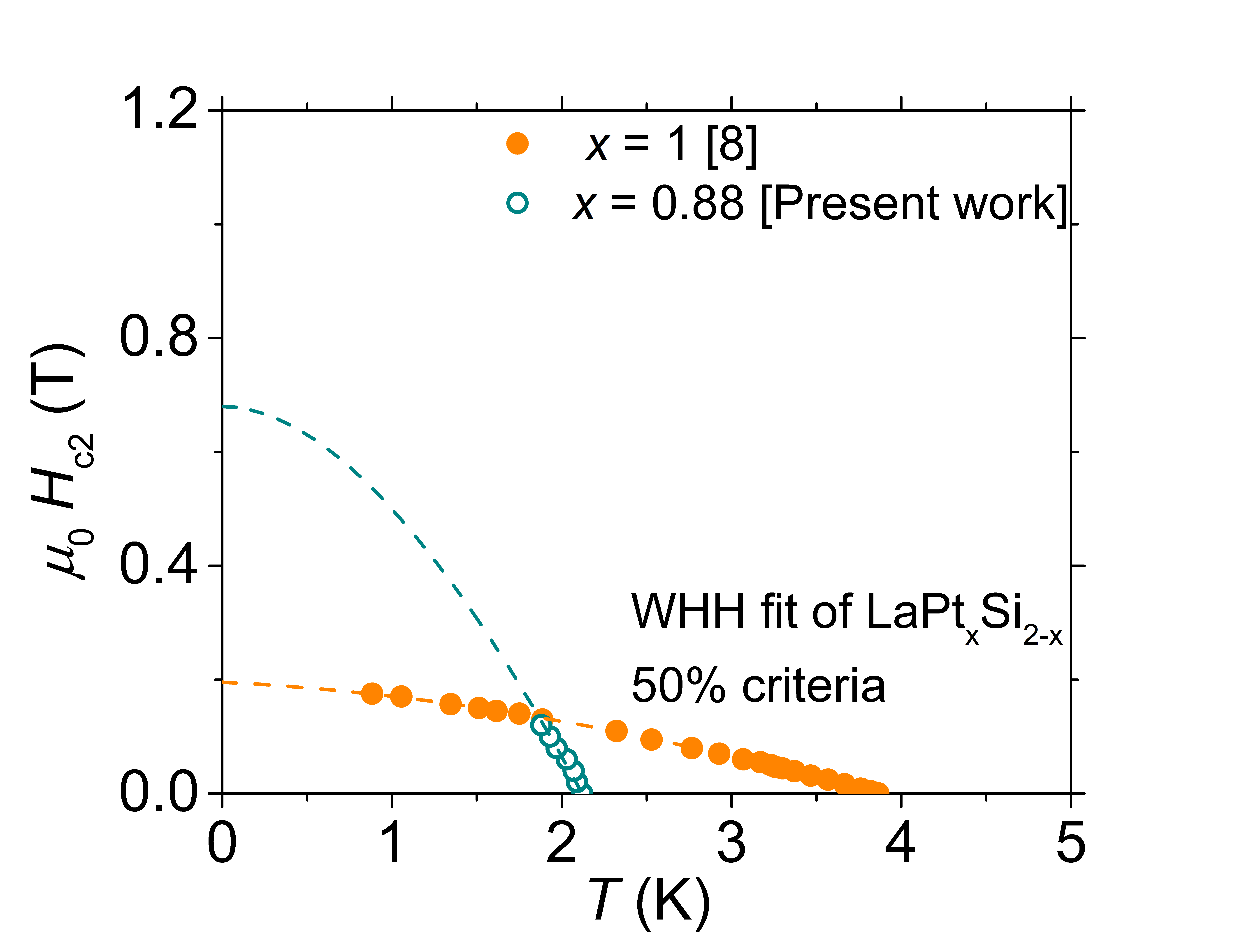}%
\caption{\label{fig:WHH} Temperature dependence of the upper critical field
$\mu_0H_{c2}$  of LaPtSi \cite{ramakrishnan1995a} and LaPt$_{0.88}$Si$_{1.12}$. The dashed
lines represent the WHH fit. One observes increased $\mu_0H_{c2}$ for the latter material.}
\end{figure}

From  Figure \ref{fig:laptsi_resi} (a), the resistivity shows the SC
$T_c$ to be 2.02 K, which
is smaller than values of 3.9 K \cite{ramakrishnan1995a}
and 3.35 K \cite{kneidinger2013a}.
The low value of residual resistivity ratio (RRR) of 1.5 compared to earlier
works \cite{ramakrishnan1995a, kneidinger2013a} can be attributed
to the disorder present in the material.
Figure \ref{fig:laptsi_resi} (b) shows the field dependent resistivity measurements where one can see the decrease
of $T_c$ with increasing magnetic fields and at 0.25 T, the superconductivity
disappears.
Upper critical fields ($\mu_0H_{c2}$) have been extracted at various temperatures using the 50\% criteria
of the resistivity drop  under different magnetic fields for
LaPt$_{0.88}$Si$_{1.12}$ as well as LaPtSi \cite{ramakrishnan1995a} to make a systematic comparison
as shown in Figure \ref{fig:WHH}.

The temperature dependence of the upper critical field is fitted by the modified form of the
Werthamer-Helfand-Hohenberg (WHH) fit as described in \cite{werthamer1966a}.
The extracted upper critical field $\mu_0H_{c2}$ at 0 K is 0.7 T (See green fit in Figure \ref{fig:WHH}) at 0 K, which is much
smaller than the Pauli paramagnetic limit 1.84 $T_c$  (3.9 T). However,
it is larger for LaPtSi, where $\mu_0H_{c2}$ = 0.2 T
(See orange fit in figure \ref{fig:WHH}) \cite{ramakrishnan1995a}.

\subsection{Magnetic susceptibility}
\begin{figure}[ht]
\centering
\includegraphics[width=80mm]{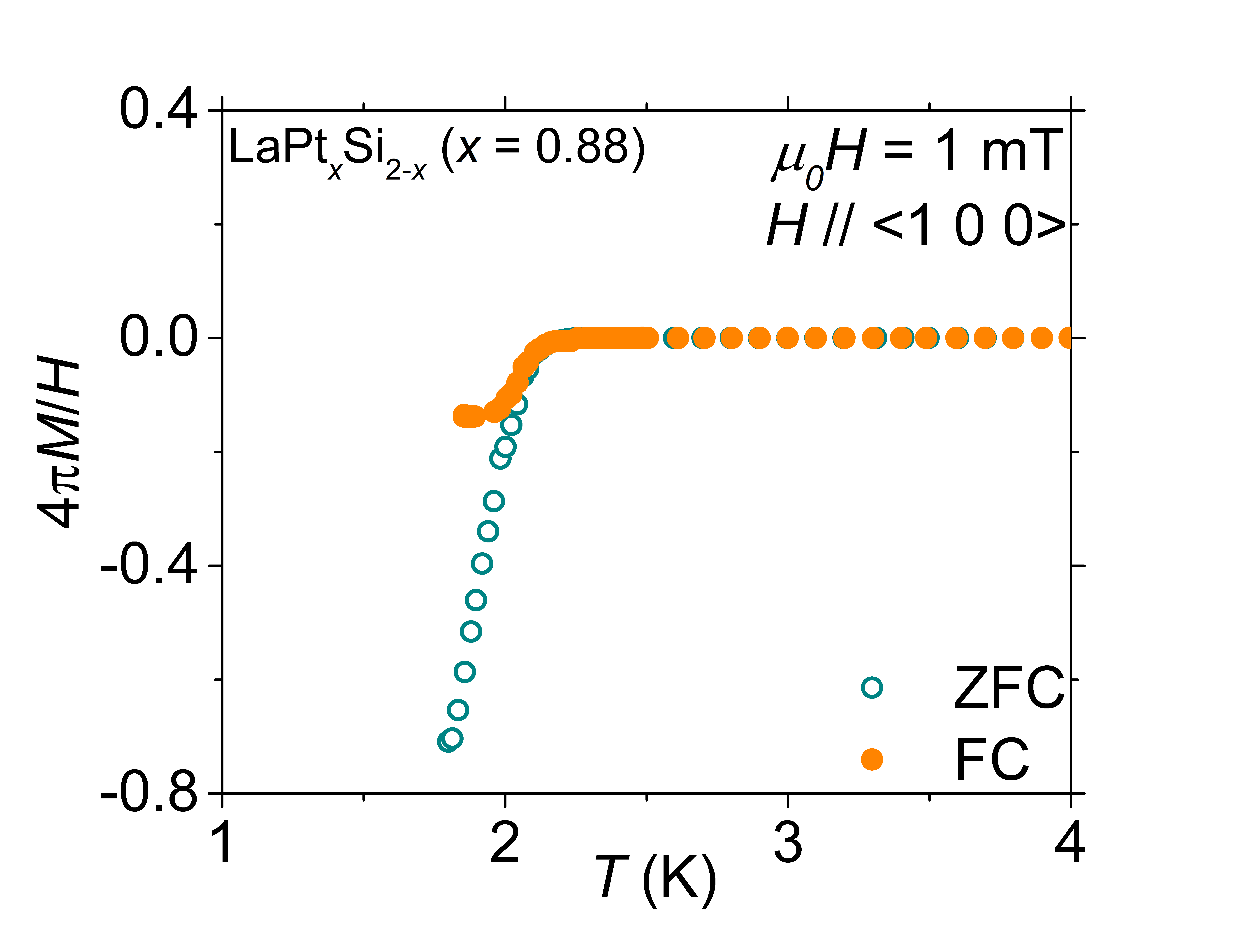}%
\caption{\label{fig:laptsi_zfc}Temperature dependence of the magnetic
susceptibility of LaPt$_{0.88}$Si$_{1.12}$ along (100) direction in a field of 1 mT in
the zero-field-cooled (ZFC) and field-cooled (FC) states.}
\end{figure}
Bulk SC is realized similar to that
of polycrystalline samples of LaPtSi \cite{ramakrishnan1995a, kneidinger2013a}.
The low temperature magnetic measurements carried out in both zero field cooled (ZFC)
and field cooled (FC) employing a small external field ($\mu_0 H$) of 1 mT.
Figure \ref{fig:laptsi_zfc}  reveals a drop in the ZFC
at 2.20 K which is in close agreement to the onset of SC as seen in the resistivity measurements.
The superconducting volume fraction of the sample is close 80\%.
In the FC data, we see a reduction in the volume susceptibility, likely
due to the trapping  of the magnetic flux.

Figure S1 in the supplemental material \cite{laptsisuppmat2023a}
shows the temperature dependence of the magnetic susceptibility
of LaPt$_{0.88}$Si$_{1.12}$ in a field of 5 T from 2 to 300 K. The normal state
susceptibility is essentially due to Pauli paramagnetism which is
temperature independent in normal metals. The rise we see below 50 K
is due to the inevitable presence of trace amounts of paramagnetic impurities in the starting
elements used in making the crystal.

\subsection{Specific heat and estimation of normal state and superconducting properties in
LaPt$_{0.88}$Si$_{1.12}$.}
\begin{figure}[ht]
\centering
\includegraphics[width=80mm]{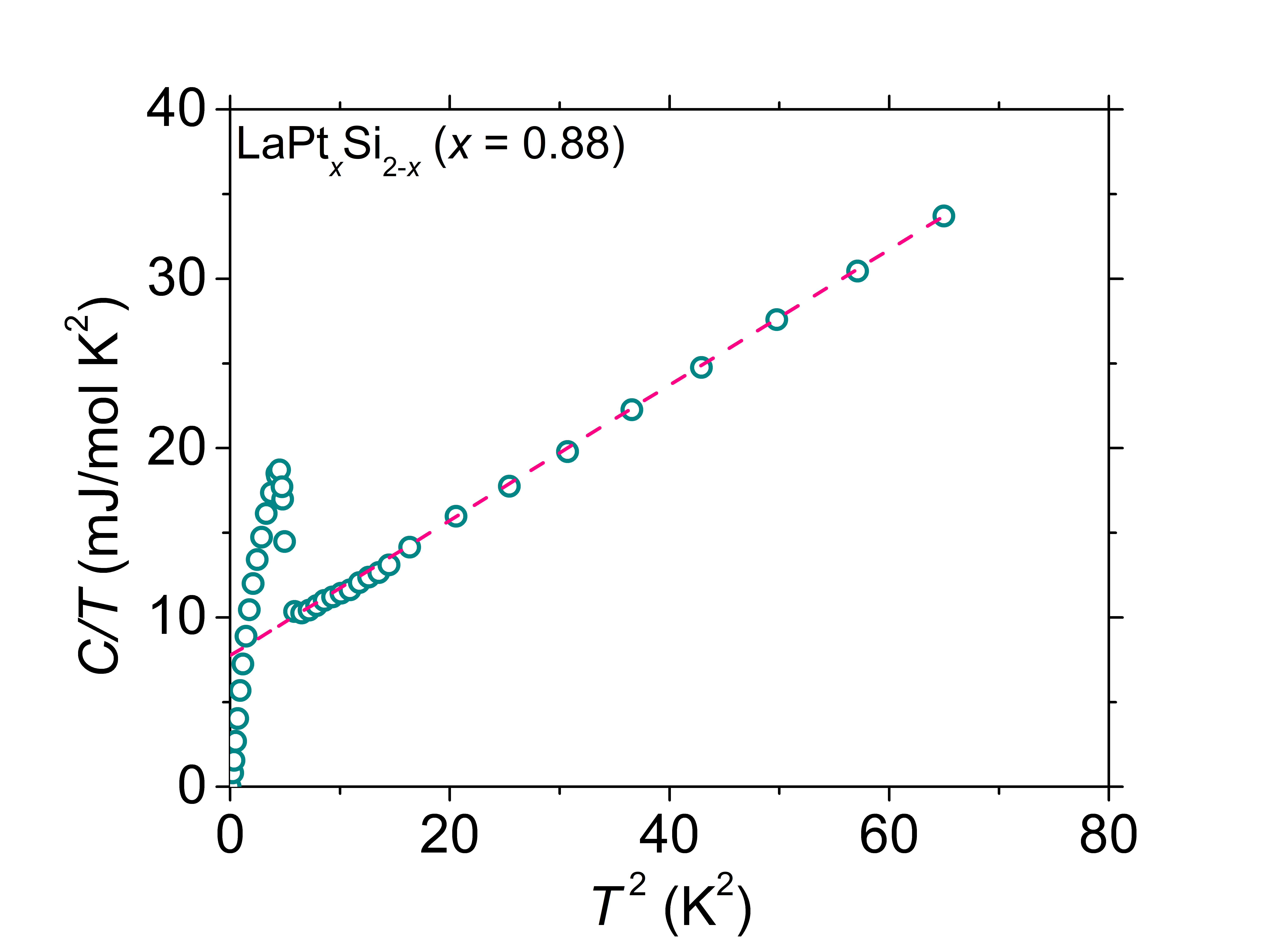}%
\caption{\label{fig:laptsi_specificheat2} Plot $C_P$/$T$ vs $T$ of LaPt$_{0.88}$Si$_{1.12}$ from 0.4 to 8 K in zero magnetic field.
The pink dashed line is the fitting curve of the normal state specific heat above 2 K.}
\end{figure}
\begin{figure}[ht]
\centering
\includegraphics[width=80mm]{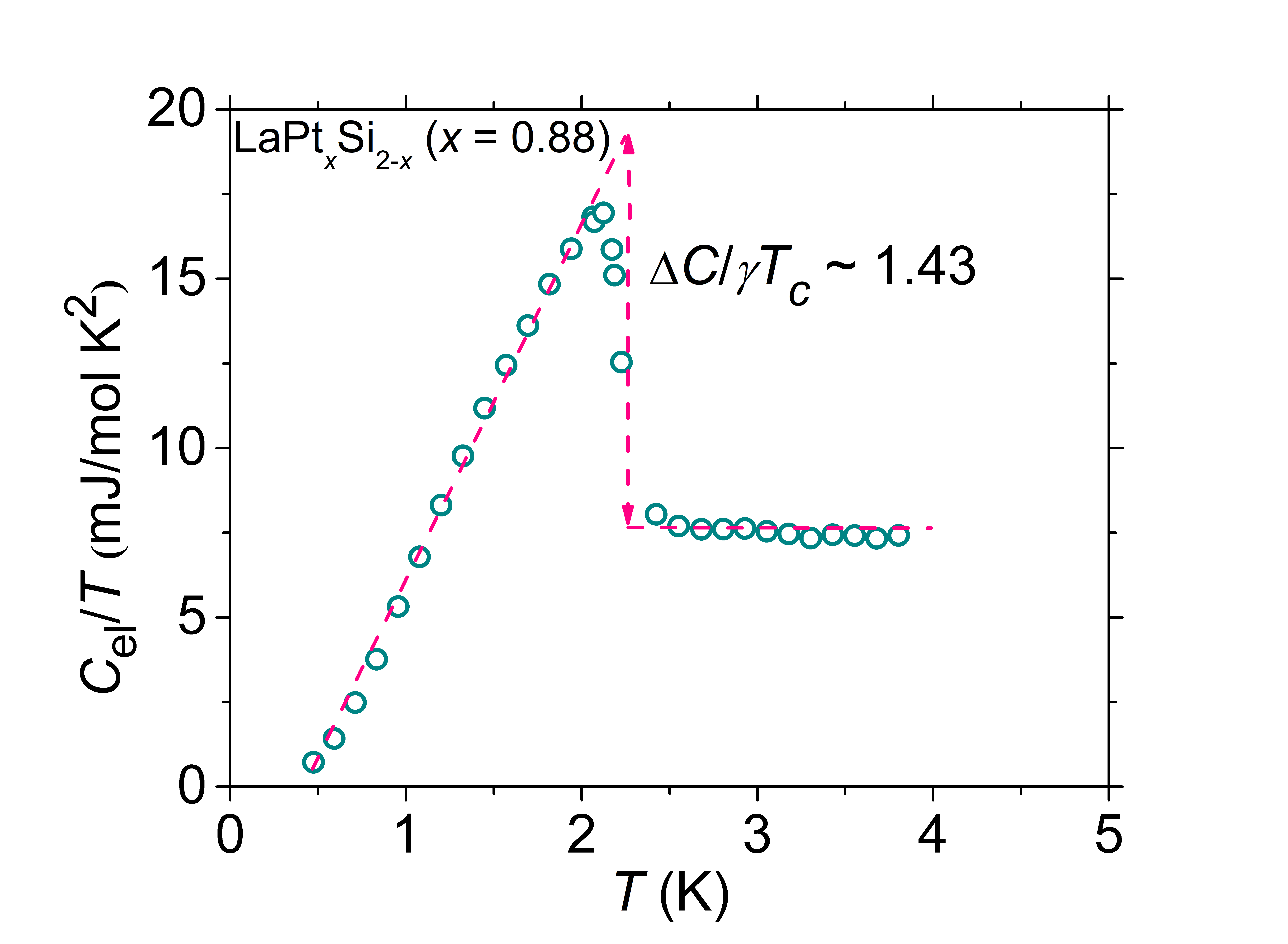}%
\caption{\label{fig:laptsi_Cel} Electronic specific heat of
LaPt$_{0.88}$Si$_{1.12}$ after the subtraction of the phonon part.
$\Delta$$C$/$\gamma$$T_c$ $\backsimeq$ 1.43 indicating that the material
is a conventional BCS superconductor. The pink dashed lines and arrow depict
 the entropy conservation process.}
\end{figure}

 Figure \ref{fig:laptsi_specificheat2} shows a jump in the specific heat at around 2.1 K which corroborates
the transition temperature determined in both magnetic susceptibility and resistivity, proving the bulk nature of superconductivity.
The data above the transition temperature can be well fitted by the formula.
\begin{equation}
C_p/T = \gamma + \beta \, T^{2}
\label{e-laptsi_1hc}
\end{equation}
where $\gamma$ is the Sommerfeld coefficient from
the electronic contribution ($\gamma$ = 7.85 mJ/mol K$^2$) and $\beta$ is the coefficient from the phonon contribution
($\beta$ = 0.40 mJ/mol K$^4$) (Figure 7a).  The Debye temperature $\Theta_D$ by using the equation
\begin{equation}
\Theta_D = (12NR\pi^{4}/5\beta)^{1/3}
\label{e-laptsi_debye}
\end{equation}
where $N$ is the number of atoms in the formula, $R$ is the gas constant. $\Theta$$_D$ is 244 K which is
slightly lower than \cite{ramakrishnan1995a} but comparable to \cite{kneidinger2013a}
In Figure \ref{fig:laptsi_Cel}, we show the electronic contribution to the heat capacity by subtracting the phonon contribution.
The superconducting jump $\Delta$$C$/$\gamma$$T_c$ is estimated to be close to 1.43 by employing the rule of entropy conservation,
indicating that it is a weakly-coupled conventional BCS superconductor \cite{bardeen1957a}.

Our experimental finding regarding that LaPt$_{0.88}$Si$_{1.12}$ a weak-coupled BCS superconductor is also in agreement
with McMillan's theory \cite{mcmillan1968a} as
the electron phonon coupling constant $\lambda$ = 0.49 calculated from the formula,
\begin{equation}
\lambda = \frac{1.04+ \mu^*ln(\Theta_D/1.45T_c)}{(1-0.62\mu^*)ln(\Theta_D/1.45T_c)-1.04}
\label{e-laptsi_lambda}
\end{equation}

The mean free path using the formula
as described by Orlando \textit{et al.} \cite{orlando1979a}
\begin{equation}
l = 1.27 \times 10^4[\rho_0(n^{2/3}S/S_F)]^{-1},
\label{e-laptsi_meanfreepath}
\end{equation}
where $n$ is the conduction electron density in units of cm$^{-3}$ and $(S/S_F)$
is the ratio of the Fermi surface to the free electron gas of density $n$. We take the ratio $(S/S_F)$
as 1 upon assumption of a spherical fermi surface. The value of $l$ is computed to be 5.83 \AA{}
which is rather small compared to the values as shown in Table \ref{comparepara}
estimated for the polycrystalline LaPtSi \cite{ramakrishnan1995a, kneidinger2013a}.

The coherence length $\xi$ is estimated to be 217 \AA{} by using the formula
\begin{equation}
\mu_{0}H_{c2}=\Phi_{0}/2\pi\xi^{2},
\label{e-laptsi_meanfreepath}
\end{equation}
As the coherence length $\xi$ is much larger than the mean free path $l$,
the system falls into the dirty limit, which is consistent with previous reports \cite{ramakrishnan1995a, kneidinger2013a}.
We note that the degree of dirtiness is more in our sample as compared to LaPtSi \cite{ramakrishnan1995a, kneidinger2013a} due the effect of Pt/Si disorder.
Table \ref{comparepara} shows the comparsion of superconducting and normal state properties
of LaPt$_{0.88}$Si$_{1.12}$ with that of LaPtSi \cite{ramakrishnan1995a, kneidinger2013a}.
\begin{table}[ht]
\scriptsize
\centering
\caption{\label{comparepara}Comparison of superconducting and normal state properties
of LaPtSi\cite{ramakrishnan1995a, kneidinger2013a} and LaPt$_{0.88}$Si$_{1.12}$.}
\centering
\begin{ruledtabular}
\begin{tabular}{cccc}
Composition & LaPtSi \cite{ramakrishnan1995a} & LaPtSi \cite{kneidinger2013a} & LaPt$_{0.88}$Si$_{1.12}$ [Present work]\\
\hline
  Type      & Polycrystal            & Polycrystal                  & Single-crystal                         \\
 $T_c$ (K)     & 3.90                 & 3.35                         & 2.02                          \\
$\rho$$_0$ ($\mu$$\Omega$cm)    & 19  & 25                          & 125                           \\
RRR                             & 24  & 9                           & 1.5                             \\
$\gamma$ (mJ/mol K$^2$)  & 3.56      & 6.50                         & 7.85                          \\
$\beta$ (mJ/mol K$^4$)   & 0.477     & 0.39                         & 0.40                            \\
$\Theta$$_D$ (K)      & 250          & 245                         & 244                             \\
$\lambda$$_{ep}$      & 0.56         & 0.54                         & 0.49                           \\
$\kappa$$_{GL}$       & 4.2         & 8.7                        & 79.2                           \\
$H_{c2}$ (T) & 0.2  & 0.4 & 0.7  \\
$\xi$ (\AA) & 338  & 283 & 217  \\
$l$ (\AA) & 43.00  & 33.07 & 5.83  \\
\end{tabular}
\end{ruledtabular}
\end{table}

From Table \ref{comparepara} we also see the calculated value of $\kappa$$_{GL}$
for LaPt$_{0.88}$Si$_{1.12}$ using the formula as described in
Orlando \textit{et al.} \cite{orlando1979a}
is 79.2, which is much higher than the values in \cite{ramakrishnan1995a, kneidinger2013a}
predominantly due to the effect of Pt/Si disorder
pushing LaPt$_{0.88}$Si$_{1.12}$ to an extreme Type-II dirty $s$-wave superconductor.

\subsection{\label{gamma}Enhanced Sommerfeld coefficient with reduction of $T_c$}

\begin{figure}[ht]
\centering
\includegraphics[width=80mm]{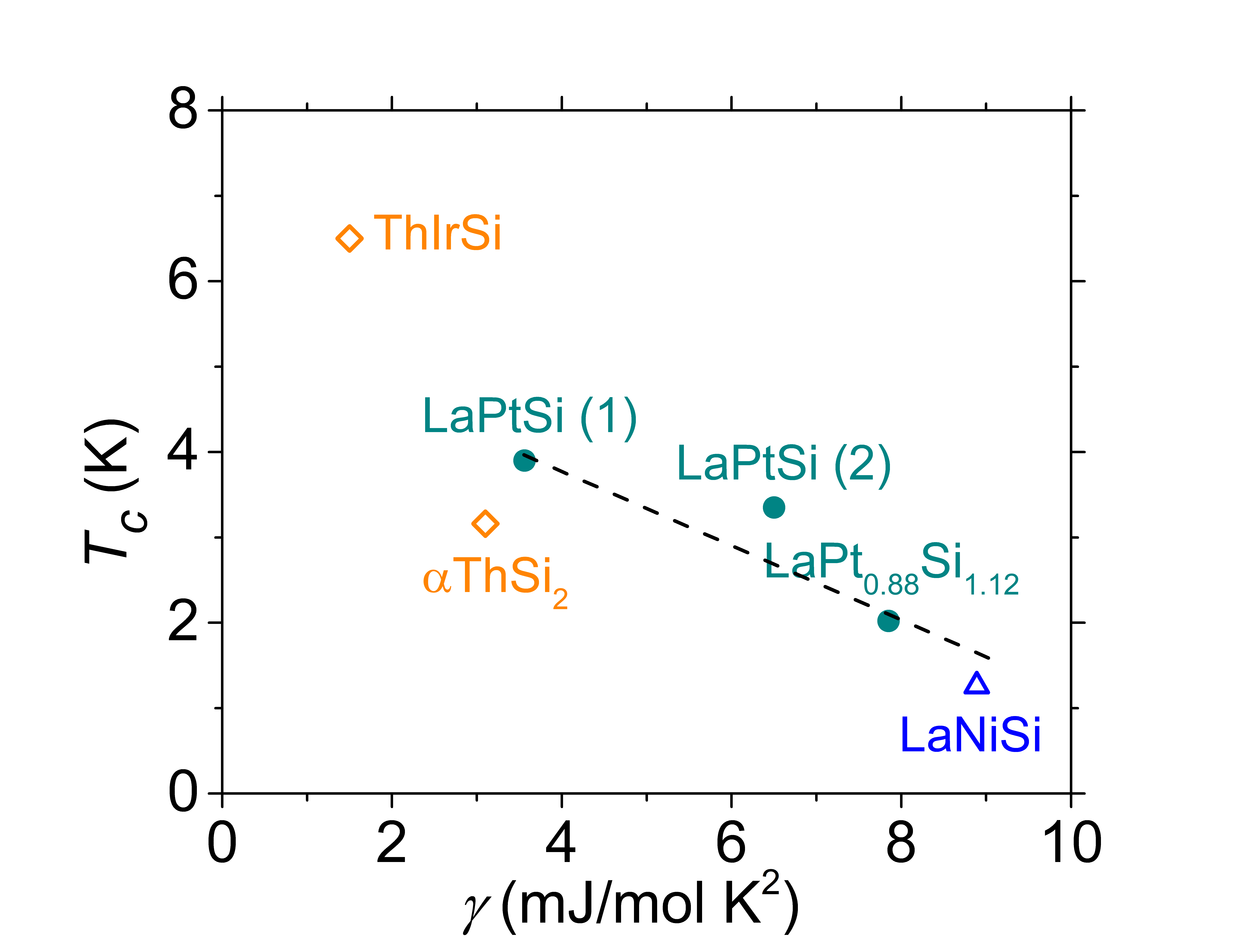}%
\caption{\label{gammatc} Variation of $T_c$ with respect
to the Sommerfeld coefficient ($\gamma$) for different systems.
One can observe that it is inversely
proportional as the $T_c$ decreases the $\gamma$ goes up. LaPtSi
(1) is from Ramakrishnan \textit{et al.}  \cite{ramakrishnan1995a}
and LaPtSi (2) is from Knedinger  \textit{et al.} \cite{kneidinger2013a}}
\end{figure}
\begin{figure}[ht]
\centering
\includegraphics[width=80mm]{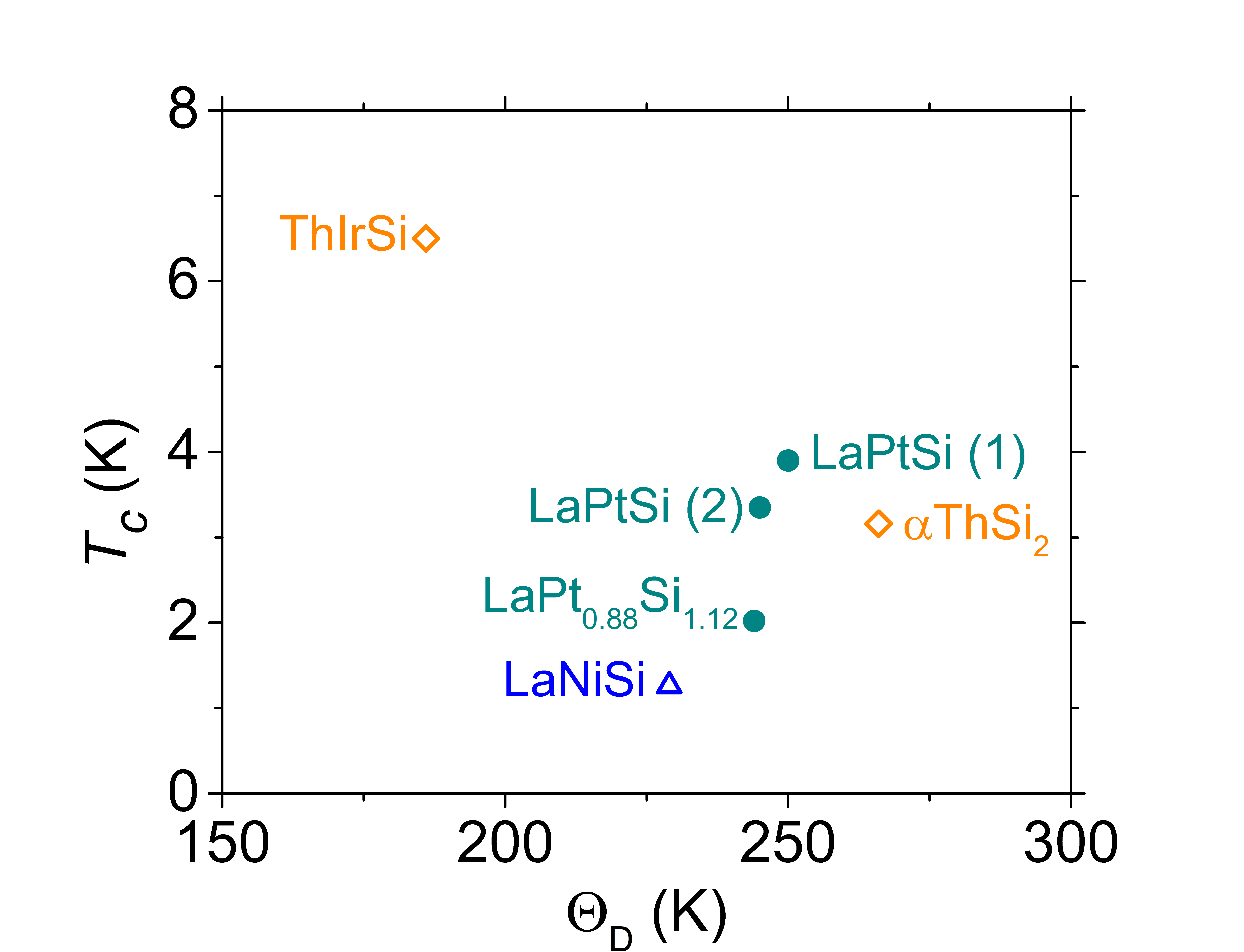}%
\caption{\label{thetatc}Variation of $T_c$ with respect
to the Debye temperature ($\Theta$$_D$).  $\Theta$$_D$ is unchanged
as the $T_c$ decreases for LaPt$_x$Si$_{2-x}$. LaPtSi
(1) is from Ramakrishnan \textit{et al.}  \cite{ramakrishnan1995a}
and LaPtSi (2) is from Knedinger  \textit{et al.} \cite{kneidinger2013a}}
\end{figure}
We observe an unusual relation between the Sommerfeld coefficient $\gamma$ and $T_c$ for LaPt$_x$Si$_{2-x}$. In many superconductors the $T_c$ increases
with $\gamma$ or density of states (DOS) at the Fermi level. However, from Figure \ref{gammatc} and Table \ref{comparepara} one can see
$T_c$ tends to be higher when $\gamma$ is smaller as indicated by LaPtSi(1),
LaPtSi(2) and LaPt$_{0.88}$Si$_{1.12}$. In another system, namely,
Cr$_{5+x}$Mo$_{35-x}$W$_{12}$Re$_{35}$Ru$_{13}$C$_{20}$ (where $x = 0{\textendash}9$) is a high entropy alloy superconductor
that shows a slight increase in $\gamma$ with reduction in $T_c$ with changes in $x$ \cite{xiao2023a} but
for LaPt$_x$Si$_{2-x}$ it is much more predominant. Figure \ref{thetatc} shows $T_c$ versus $\Theta$$_D$ for LaPt$_x$Si$_{2-x}$
and other isostructural compounds. $\Theta$$_D$ is almost unchanged while $T_c$ is reduced from 3.9 to 2.02 K with increasing $x$
in LaPt$_x$Si$_{2-x}$. These observations suggest that the $T_c$ is determined by the electronic properties and not by phonons
for LaPt$_x$Si$_{2-x}$.

\subsection{Sharp fall of superconductivity in LaPt$_{x}$Si$_{2-x}$ compared to other isostructural compounds}
\begin{figure}[ht]
\centering
\includegraphics[width=80mm]{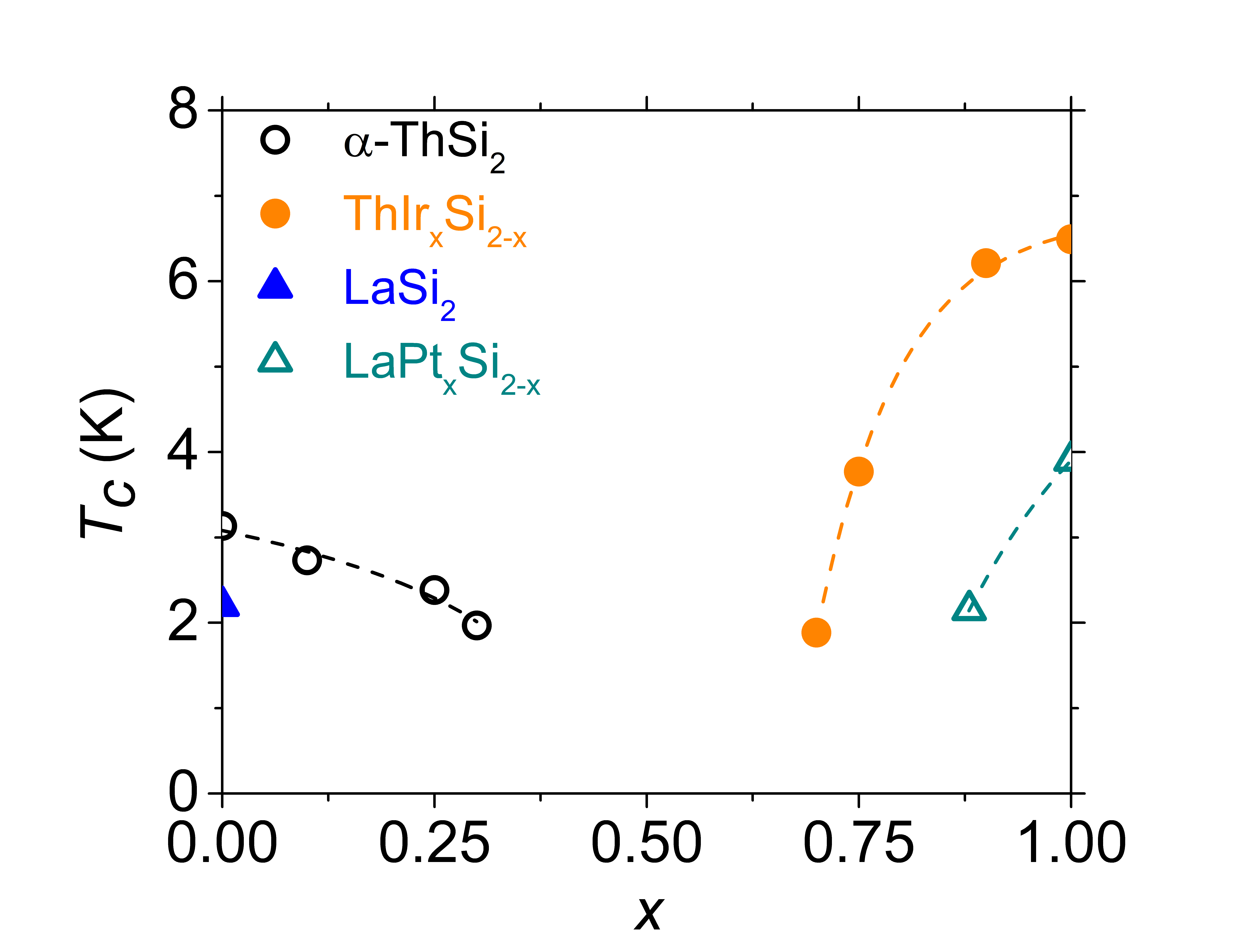}%
\caption{\label{fig:laptsi_scdome}The decrease of the superconducting transition
temperature with respect to $x$ for compounds $\alpha$-ThSi$_2$ \cite{leejay1983a},
LaSi$_2$ \cite{satoh1970a}, ThIr$_x$Si$_{2-x}$ \cite{leejay1983a}
and LaPt$_{x}$Si$_{2-x}$ \cite{ramakrishnan1995a} where
one observes that the latter compound exhibits much sharper
 reduction in $T_c$ with decrease in $x$.}
\end{figure}
Figure \ref{fig:laptsi_scdome} shows the change in $T_c$ with
respect to the $x$ concentration in a few compounds that are structurally similar
to LaPt$_{x}$Si$_{2-x}$. The left side of the plot
depicts the centrosymmetric superconductors $\alpha$-ThSi$_2$ and LaSi$_2$, while on
the right side we shown the non-centrosymmetric superconductors ThIrSi and LaPtSi.
Combining the results of LaPt$_{0.88}$Si$_{1.12}$ and LaPtSi, the suppression of
superconductivity for LaPt$_{x}$Si$_{2-x}$ is much  sharper  in comparison to ThIr$_x$Si$_{2-x}$
and also another isostructural compound LaNi$_x$Pt$_{1-x}$ \cite{lee1995a}.
From Figure \ref{fig:laptsi_scdome} based on the trajectory of the fall of SC in LaPt$_{x}$Si$_{2-x}$ one can expect a
a complete suppression of the $T_c$ when $x$ reaches close to 0.75.

Another interesting point is that the polycrystalline sample grown by Kneidinger \textit{et al.} \cite{kneidinger2013a} exhibits a $T_c$ at 3.35 K, an intermediate value
lying in between $T_{cs}$ = 2.02 K and 3.9 K for the samples LaPt$_{0.88}$Si$_{1.12}$ (present work) and LaPtSi ( Ramakrishnan \textit{et al.}  \cite{ramakrishnan1995a}).
We have discussed that the decrease in $x$  in LaPt$_{x}$Si$_{2-x}$ leads to a sharp drop in $T_c$ which leads us to speculate
on the reason behind the lower value of $T_c$ reported by Kneidinger \textit{et al.} as compared to Ramakrishnan \textit{et al.}. It is possible
that $x$ in Kneidinger \textit{et al.}'s sample could be much higher than that of our single-crystal ($x = 0.88$) but less than 1. However,
such a small decrease in $x$  from 1 may be difficult to detect from PXRD measurements.
This could be a plausible reason as to why it was reported as stoichiometric LaPtSi by Kneidinger \textit{et al.}. From Table \ref{comparepara}
we also infer that $\gamma$ of Kneidinger \textit{et al.}'s sample is 6.50 mJ/mol K$^2$ sandwiched between the values reported by Ramakrishnan \textit{et al.}
and the present work showing an enhancement of $T_c$ with decrease in $\gamma$ as discussed in Section III E.
Regardless of this hypothesis, it can be evidently seen that the reduction of $T_c$ in LaPt$_{x}$Si$_{2-x}$ as $x$ decreases
is very steep compared to ThIr$_x$Si$_{2-x}$  \cite{leejay1983a, chevalier1986a} and LaNi$_x$Pt$_{1-x}$Si \cite{lee1995a}.

\section{\label{sec:Laptsi_conclusions}Conclusions}
To summarize, we find that site disordered LaPt$_{0.88}$Si$_{1.12}$ crystal shows superconductivity around 2 K as
observed by resistivity, magnetic susceptibility and heat capacity measurements. Analysis
of the results suggest that the superconductivity arises due to  weak electron-phonon coupling as illustrated by the BCS theory.
However, in the disordered material the SC transition is significantly reduced as compared to those of  stoichiometric polycrystalline samples of LaPtSi.
It is pertinent to state here that the reduction in $T_c$ is more than that observed in other similar systems, which warrants further investigation. In addition, the Sommerfeld
coefficient shows a two-fold enhancement even though $T_c$ is decreased to 2.02 K. The extrapolated value of the upper critical field at
0 K is much larger than the values quoted for the polycrystalline samples. We propose that a rapid reduction of $T_c$
with concomitant increase in $\gamma$
 with $x$ in LaPt$_x$Si$_{2-x}$ suggests underlying non-trivial topological bands that could
be responsible for the observed unusual disorder effects on the SC and normal state properties. In order to understand further,
one must  study the Fermi surface of the present crystal using angle resolved photoemission (ARPES) studies to look
for  an interplay of the disorder, Dirac points, and superconductivity.
\begin{acknowledgments}
Single-crystal x-ray diffraction data was collected
at SAIF laboratory, IIT Madras, India. We thank Mr.
Y. Shibata for EPMA measurements at Hiroshima University. Parts of this
research was supported by JSPS KAKENHI Grant Nos: JP22K03529, JP21K03448, J23H04630, JP23H04870,
JGC-S Scholarship Foundation No: 2010,
The Hattori Hokokai Foundation No: 21-010 and
The Mazda Foundation No: 21KK-191.

\end{acknowledgments}

%

\end{document}